\documentclass[iop]{emulateapj}

\shorttitle{RV~Tau stars in Gaia DR2}
\shortauthors{B\'odi \& Kiss}

\usepackage{amsmath}

\begin{document}

\title{Physical properties of galactic RV Tauri stars from Gaia DR2 data}

\author{A. B\'odi\altaffilmark{1,2} and L.L. Kiss\altaffilmark{1,3}}
\email{bodi.attila@csfk.mta.hu}

\altaffiltext{1}{Konkoly Observatory, Research Centre for Astronomy and Earth Sciences, Hungarian Academy of Sciences, H-1121 Budapest, Konkoly Thege M. \'ut 15-17, Hungary}
\altaffiltext{2}{MTA CSFK Lend\"ulet Near-Field Cosmology Research Group}
\altaffiltext{3}{Sydney Institute for Astronomy, School of Physics A29, University of Sydney, NSW 2006, Australia}

\begin{abstract}
We present the first period-luminosity and period-radius relation of Galactic RV Tauri variable stars. We have surveyed the literature for all variable stars belonging to this class and compiled the full set of their photometric and spectroscopic measurements. We cross-matched the final list of stars with the Gaia DR2 database and took the parallaxes, G-band magnitudes and effective temperatures to calculate the distances, luminosities and radii using a probabilistic approach. As it turned out, the sample was very contaminated and thus we restricted our study to those objects for which the RV Tau-nature was securely confirmed. We found that several stars are located outside the red edge of the classical instability strip, which implies a wider pulsational region for RV Tau stars. The period-luminosity relation of galactic RV Tauri stars is steeper than that of the shorter-period Type II Cepheids, in agreement with previous result obtained for the Magellanic Clouds and globular clusters. The median masses of RVa and RVb stars were calculated to be 0.45-0.52 M$_{\odot}$ and 0.83 M$_{\odot}$, respectively.

\end{abstract}


\section{Introduction}

RV Tauri-type variables form the long-period extension of Population II Cepheids, which are metal-poor low- and intermediate mass F-, G- and K-type supergiant stars, older than the classical Cepheids (see \citet{Wallerstein02} for a review). RV Tauris are among the most luminous stars (10$^3$ - 10$^4$ L$_\odot$), having already left the red giant (RGB) or the asymptotic giants branch (AGB) and rapidly evolving through the post-RGB/AGB instability strip to become planetary nebulae \citep{Manick18,Kamath15,Kamath14,Jura86}. Hence, they provide important information about the not so well-known late-phases of stellar evolution, in which pulsations and mass-loss processes can interact and influence stellar evolution. 

During their evolution, post-AGB stars cross the instability strip where they become unstable against radial pulsation. The observed periods of RV Tau stars usually fall between 20 and 90 days \citep{Soszynski10,Soszynski08}. The main characteristics of the light curve is the presence of alternating minima (i.e. every second minima are shallower). The periodicity is not strict, as the cycle-to-cycle variations can be quite significant, and in some cases, it has been shown to be caused by low-dimensional chaos (e.g. \citealt{Buchler96}, \citealt{Plachy18}). In addition to the Cepheid-like pulsations, some RV Tau stars show long-term modulation of the mean brightness with periods of 700-2500 days, associated with time-variable dust obscuration \citep{Kiss17}. The absence or presence of the slow modulation is the basis for classifying the stars into the RVa and RVb photometric sub-classes, respectively.

Previous studies on the period-luminosity (PL) relations of RV Tauri stars were almost exclusively based on various samples of Population II Cepheids in the Magellanic Clouds or globular clusters. There were hints of a different slope for longer-period Type II Cepheids \citep{McNamara95}, which was found to depend on the wavelength of the 
observations, with negligible effects in the JHK$_{\rm S}$ bands \citep{Matsunaga06}. Recently, \citet{Groenewegen17} and \citet{Manick17} presented supporting indications of a steeper RV Tau PL relation for the dusty objects in the Magellanic Clouds and the Milky Way, respectively. 

Until now no PL relation has been published for nearby, bright and in all other aspects well-observed Galactic RV Tauri stars. Gaia DR2 has opened, for the first time, the possibility of a geometric distance measurement of Galactic RV Tau-stars. The main inspiration of our study is to compare the RV Tau populations of the Milky Way and the Magellanic Clouds and the universality of their PL relations.  

\section{Data and methods}

To identify all the known stars in the Galaxy that are thought to be of RV Tau-type, we searched the catalog of variable star index (VSX database\footnote{http://www.aavso.org/vsx/}), the General Catalog of Variable Stars (GCVS; \citealt{Samus17}) and the SIMBAD \citep{Wenger00} database. Then, we cross-correlated this sample with the Gaia archive\footnote{https://gea.esac.esa.int/archive/} \citep{Gaia16,Gaia18} and downloaded all available measurements for stars that have relative parallax error ($\sigma_\pi/\pi$)  smaller than 0.2 (well-determined parallaxes; \citealt{Astraatmadja16}). We have found 56 stars. We note that in most cases extinction values are missing in the Gaia table. This turned out to be a consequence of the data filtering done by the Gaia team (see Sect. 6.5, Eqs. (8-11) and Figs. 31 and 19d in \citealt{Andrae18}), which practically removed all stars between the upper main sequence and the RGB/AGB in the Hertzsprung--Russell-diagram. Because of this and the strong degeneracy between extinction and temperatures, the Gaia T$_{\rm eff}$ values were also found to be systematically biased (see details below). 

\citet{Bailer-Jones15} showed that distance estimation from parallaxes becomes an inference problem when measurement errors are present. Traditional inverse approach gives an incorrect (symmetric) error estimate, which can be avoided by using a properly normalized prior. \citet{Astraatmadja16} investigated the performance of various priors for estimating distances and found that the exponentially decreasing space density (EDSD) prior performs well with a length scale of 1.35 kpc. To determine Gaia distances we followed  the prescription of the EDSD method.

To calculate the absolute magnitudes and luminosities, we used photometric measurements, extinction and bolometric correction values. 2MASS J,H,K$_s$ and Johnson V,I band photometric values were taken from the SIMBAD catalog. However, during the calculation of luminosities, the Gaia magnitudes were preferred, if available.
Extinctions A$_{\rm V}$ were taken from the combination of 3D reddening maps by \citet{Marshall06}, \citet{Green15}, and \citet{Drimmel03} as implemented in the python package of {\tt mwdust} \citep{Bovy16}. Bolometric corrections (BC) were linearly interpolated from T$_{\rm eff}$ , log$g$, [Fe/H] and A$_{\rm V}$ in the MIST/C3K grid (C. Conroy et al., in preparation\footnote{http://waps.cfa.harvard.edu/MIST/model\_grids.html}). T$_{\rm eff}$ , log$g$ and [Fe/H] values were taken from the literature or the Gaia archive. Missing log$g$ and [Fe/H] values were filled by 0.0, assuming the errors to be 0.4 and 0.5, respectively. These parameters have subtle influence on the BCs.

Absolute magnitude, luminosity and radius values were determined in a probabilistic approach using the direct mode of the slightly modified {\tt isoclassify} code of \citet{Huber17}, which uses a Monte-Carlo sampling scheme and derives posterior distributions of all parameters. The parameter estimation was performed star by star as follows. First, the distance is determined, then the A$_{\rm V}$ is estimated from the reddening map, which is used to calculate the absolute magnitude in an available photometric band. Using BC, luminosity is determined, which is then converted into radius using the effective temperature and the Stefan-Boltzmann law. The final values and the errors are the median and the 1$\sigma$ confidence interval of the distributions.

The periods of pulsation and mean-brightness variation (in case of the RVb stars) were taken from the literature. For stars without published periods, we downloaded the AAVSO (American Association
of Variable Star Observers) or ASAS (All Sky Automated Survey; \citealt{Pojmanski02}) light curves and determined the pulsation periods from the Fourier-spectra, if it was possible. In this paper, we consistently use the double-periods, i.e. the duration between two executive shallow or deep minima as the length of the pulsation cycles.

\subsection{Sample definition}

   \begin{figure}
   \centering
   \includegraphics[width=9cm]{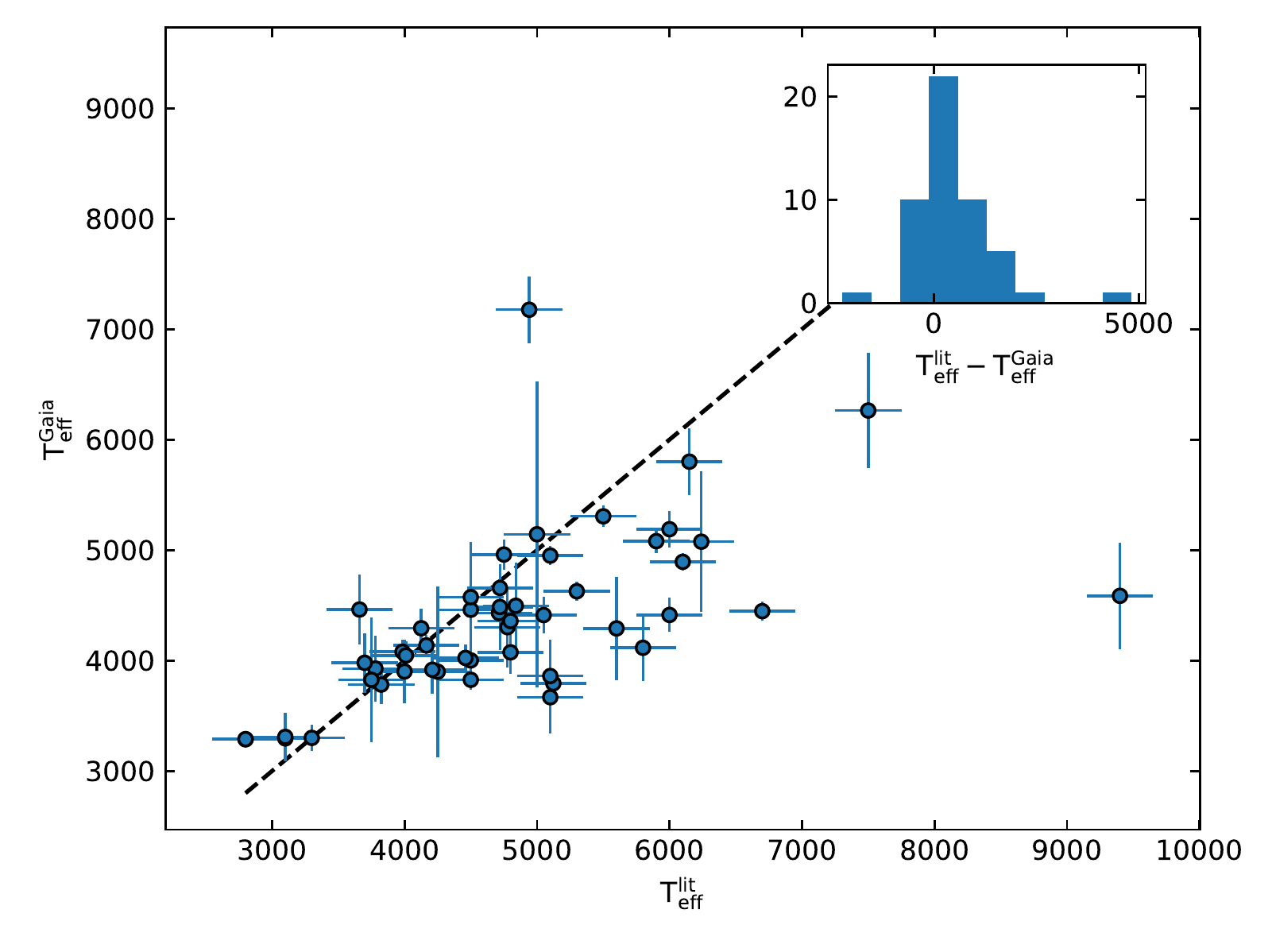}
      \caption{A comparison of effective temperatures obtained from Gaia and those derived using spectroscopy. The black dashed line shows the 1:1 relation. The increasing deviation is clearly visible as the literature T$_{\rm eff}$ increases. The insert shows the distribution of the differences.}
         \label{teff_dist}
   \end{figure}

When we plotted the Hertzsprung--Russell-diagram and the period-luminosity relation, we found that the observational scatter was huge. The initial sample contained several low-luminosity stars, and also objects located far from the theoretical instability strip. We interpreted the large scatter as due to contamination of misclassified objects, hence we performed a strict revision of the sample as follows.

First, we thoroughly surveyed the literature for all stars in the initial sample to remove all objects for which there was the slightest doubt about the RV Tau-nature. Second, we took into account the systematic investigation of misclassified RV Tauris by \citet{Zsoldos91}. Third, we also excluded stars with poorly determined pulsation periods (meaning that we may have left out genuine RV Tauris, too, which need further photometric observations to measure the periods).  

In the next step, we reviewed the AAVSO and ASAS light curves for each star remaining and checked the variability characteristics by visual inspection. Stars with few observations and those that do not show the alternating behavior clearly (i.e. every minima have almost the same depth) were filtered out. During the inspection, we found incorrectly determined magnitudes in the commonly used catalogues.

RVb-type stars needed a specific treatment in terms of calculating their luminosities. As it turned out, the catalogued mean magnitude values were previously determined by averaging the brightness over the whole light curve, including the long-term RVb cycles. However, given that the RVb phenomenon is indeed connected to the dust extinction around the stars \citep{Kiss17}, we re-determined the mean brightnesses of these objects by averaging the light curves only in the vicinity of the maximum of the long-term variation. This resulted in much higher luminosities, which in turn decreased the scatter of the PL relation.

Finally, we decided to restrict our detailed analysis to those stars that: (i) were included in the spectral energy distribution (SED) study of \citet{Gezer15} as objects that were chemically studied before; and (ii) the RVb stars of \citet{Kiss17} that have spectroscopically determined parameters. This is the most reliable, high confidence collection of galactic RV Tauri-type variables with well-determined Gaia DR2 distances, which contains 12 RVa- and 6 RVb-type stars in the Galaxy. 

\begin{table*}
\centering
\caption{The physical parameters of high-confidence galactic RV Tauri stars (see text for details). The errors represent the 1$\sigma$ confidence level of the posterior distributions. The effective temperatures were taken predominantly from \citet{Gezer15} and are all based on spectroscopic measurements. Periods of pulsation and mean-brightness variation were calculated by us or were taken from the literature.}
\begin{tabular}{ccccccccccc}
\hline \hline
Name & d [pc] & L/L$_{\odot}$ & R/R$_{\odot}$ & M$_{\rm V}$ [mag] & T$_{\rm eff}$ [K] & Period [d] & RVb period [d] & Ref. & $\pi$ [mas] & $\sigma_\pi$ [mas]\\
\hline
\multicolumn{8}{c}{RVa type stars} \\
\hline
UY Ara & 4648$^{+1064}_{-629}$ & 884$^{+510}_{-204}$ & 34.4$^{+8.2}_{-5.5}$ & $-$2.730$^{+0.324}_{-0.464}$ & 5500 & 56.94 & \nodata & 1 & 0.21 & 0.04 \\
EQ Cas & 4336$^{+802}_{-534}$ & 844$^{+361}_{-164}$ & 48.4$^{+11.7}_{-7.3}$ & $-$2.460$^{+0.286}_{-0.367}$ & 4500 & 58.23 & \nodata & 2 & 0.22 & 0.03 \\
RU Cen & 1932$^{+221}_{-158}$ & 1054$^{+304}_{-169}$ & 31.1$^{+4.9}_{-3.5}$ & $-$3.191$^{+0.198}_{-0.226}$ & 6000 & 64.74 & \nodata & 3 & 0.52 & 0.05 \\
V820 Cen & 2260$^{+397}_{-248}$ & 1964$^{+915}_{-407}$ & 68.0$^{+14.7}_{-11.0}$ & $-$3.004$^{+0.263}_{-0.368}$ & 4750 & 159.70  & \nodata & 1 & 0.45 & 0.06 \\
SS Gem & 3423$^{+836}_{-488}$ & 17680$^{+12800}_{-6400}$ & 150.6$^{+41.7}_{-34.8}$ & $-$6.685$^{+0.588}_{-0.523}$ & 5600 & 89.83 & \nodata & 1 & 0.29 & 0.05 \\
AC Her & 1276$^{+49}_{-44}$ & 2475$^{+183}_{-209}$ & 47.1$^{+4.7}_{-4.1}$ & $-$3.929$^{+0.084}_{-0.084}$ & 5900 & 75.46 & \nodata & 3 & 0.78 & 0.03 \\
EP Lyr & 6170$^{+1267}_{-806}$ & 4164$^{+2075}_{-902}$ & 61.2$^{+13.9}_{-9.3}$ & $-$4.845$^{+0.299}_{-0.410}$ & 6100 & 83.18 & \nodata & 2 & 0.15 & 0.03 \\
TT Oph & 2535$^{+221}_{-172}$ & 714$^{+131}_{-102}$ & 38.5$^{+5.4}_{-4.5}$ & $-$2.791$^{+0.165}_{-0.165}$ & 4800 & 61.08 & \nodata & 3 & 0.40 & 0.03 \\
R Sge & 2475$^{+353}_{-229}$ & 2329$^{+744}_{-638}$ & 61.2$^{+12.4}_{-9.9}$ & $-$3.505$^{+0.302}_{-0.302}$ & 5100 & 71.15 & \nodata & 1 & 0.40 & 0.05 \\
AR Sgr & 2823$^{+516}_{-258}$ & 1878$^{+756}_{-412}$ & 53.5$^{+10.8}_{-8.4}$ & $-$3.521$^{+0.252}_{-0.353}$ & 5300 & 86.76 & \nodata & 1 & 0.34 & 0.05 \\
V Vul & 1854$^{+160}_{-140}$ & 2169$^{+504}_{-315}$ & 77.9$^{+13.0}_{-10.1}$ & $-$3.873$^{+0.203}_{-0.237}$ & 4500 & 76.09 & \nodata & 1 & 0.55 & 0.04 \\
\hline
\multicolumn{8}{c}{RVb type stars} \\
\hline
TW-Cam & 1771$^{+142}_{-106}$ & 2157$^{+386}_{-257}$ & 68.0$^{+9.5}_{-8.5}$ & -3.361$^{+0.168}_{-0.168}$ & 4800 & 86.36 & 671 & 6 & 0.57 & 0.04 \\
IW Car & 1907$^{+128}_{-96}$ & 2622$^{+338}_{-296}$ & 37.9$^{+4.0}_{-3.5}$ & $-$3.867$^{+0.127}_{-0.127}$ & 6700 & 71.98 & 1449 & 4 & 0.52 & 0.03 \\
SX Cen & 4429$^{+1071}_{-605}$ & 3684$^{+2315}_{-842}$ & 61.1$^{+14.7}_{-9.8}$ & $-$4.343$^{+0.319}_{-0.479}$ & 6000 & 32.88 & 602 & 4 & 0.22 & 0.04 \\
DF Cyg & 2737$^{+240}_{-186}$ & 815$^{+155}_{-116}$ & 39.9$^{+6.4}_{-4.5}$ & $-$2.286$^{+0.189}_{-0.189}$ & 4840 & 49.82 & 780 & 4 & 0.37 & 0.03 \\
BT Lac & 3034$^{+307}_{-204}$ & 686$^{+117}_{-192}$ & 32.3$^{+5.3}_{-4.5}$ & $-$1.436$^{+0.417}_{-0.179}$ & 5050 & 40.78 & 650 & 2,5 & 0.33 & 0.03 \\
U Mon & 1111$^{+137}_{-102}$ & 5480$^{+1764}_{-882}$ & 100.3$^{+18.9}_{-13.2}$ & $-$4.516$^{+0.246}_{-0.287}$ & 5000 & 91.48 & 2451 & 4 & 0.92 & 0.09 \\
RV Tau & 1460$^{+153}_{-117}$ & 2453$^{+605}_{-403}$ & 83.4$^{+12.8}_{-12.8}$ & $-$3.359$^{+0.213}_{-0.244}$ & 4500 & 78.48 & 1210 & 4 & 0.69 & 0.06 \\
\hline
\end{tabular}
\label{tab1}
\tablerefs{ (1) \citet{Wils08}; (2) this paper; (3) \citet{Samus17}; (4) \citet{Kiss17}; (5) \citet{Percy15}; (6) \citet{Manick17} }
\end{table*}

\subsection{Gaia versus spectroscopic T$_{\rm eff}$ values}

The second Gaia data release (Gaia DR2) contains photometry in three different bands. G is a broad band, while BP and RP were obtained by integrating the red and blue side of the grism spectra. From the difference of these Gaia bands, stellar effective temperatures were inferred for stars brighter than G=17 mag with T$_{\rm eff}$ between 3000-10000 K. These results are reliable within an accuracy of 324 K \citep{Andrae18}. This value represents the random errors, not taking into account the systematic uncertainties.

As we found spectroscopically determined (hence expected to be more reliable in relation to interstellar reddening) effective temperatures for several RV Tauri type stars, we can compare the Gaia inferred T$_{\rm eff}$ values with the literature to test the reliability of the given error bars. Fig. \ref{teff_dist} shows the T$^{\rm Gaia}_{\rm eff}$ vs. T$^{\rm lit}_{\rm eff}$ overplotted with the distribution of differences in insert. As can be seen in the plot, the temperatures are in reasonable agreement within the given error bars below T$^{\rm lit}_{\rm eff} \sim$ 4500~K. After this point, the deviation increases with increasing T$^{\rm lit}_{\rm eff}$ (except one point which is covered by the insert). This increasing deviation is expected from Fig. 11.(c) of \citet{Andrae18}, but this should be symmetric. From the distribution in insert, we can estimate a mean deviation of 445.4 K, which significantly decreases the accuracy of the Gaia temperatures. This effect strongly influences the position of the RV Tau stars in the HR diagram (next subsection) and the mass estimation (Section \ref{sect_mass}), if no spectroscopic effective temperatures are available. That is why we restricted
out investigation to those stars for which spectroscopic temperatures were available from the literature.

As has been pointed out by the referee, Gaia parallaxes are based on single star solution while many of the RV Tauris (especially the RVbs and the disk stars) are found in binaries that might affect the observed discrepancy between the Gaia and spectroscopic effective temperatures. However, \citet{Manick17} showed that the spectra of RVb stars are dominated by the highly luminous primary star and no signature of the companion is seen in the spectra. Thus, we expect that the influence of the secondary on the determination of the effective temperatures is negligible.

The finally adopted fundamental physical parameters are listed in Table \ref{tab1}. These form the basis of the detailed discussion in the next Section.

\section{Results}

\subsection{The empirical Hertzsprung--Russell-diagram}
\label{sect:HRD}

\begin{figure}
   \centering
   \includegraphics[width=9cm]{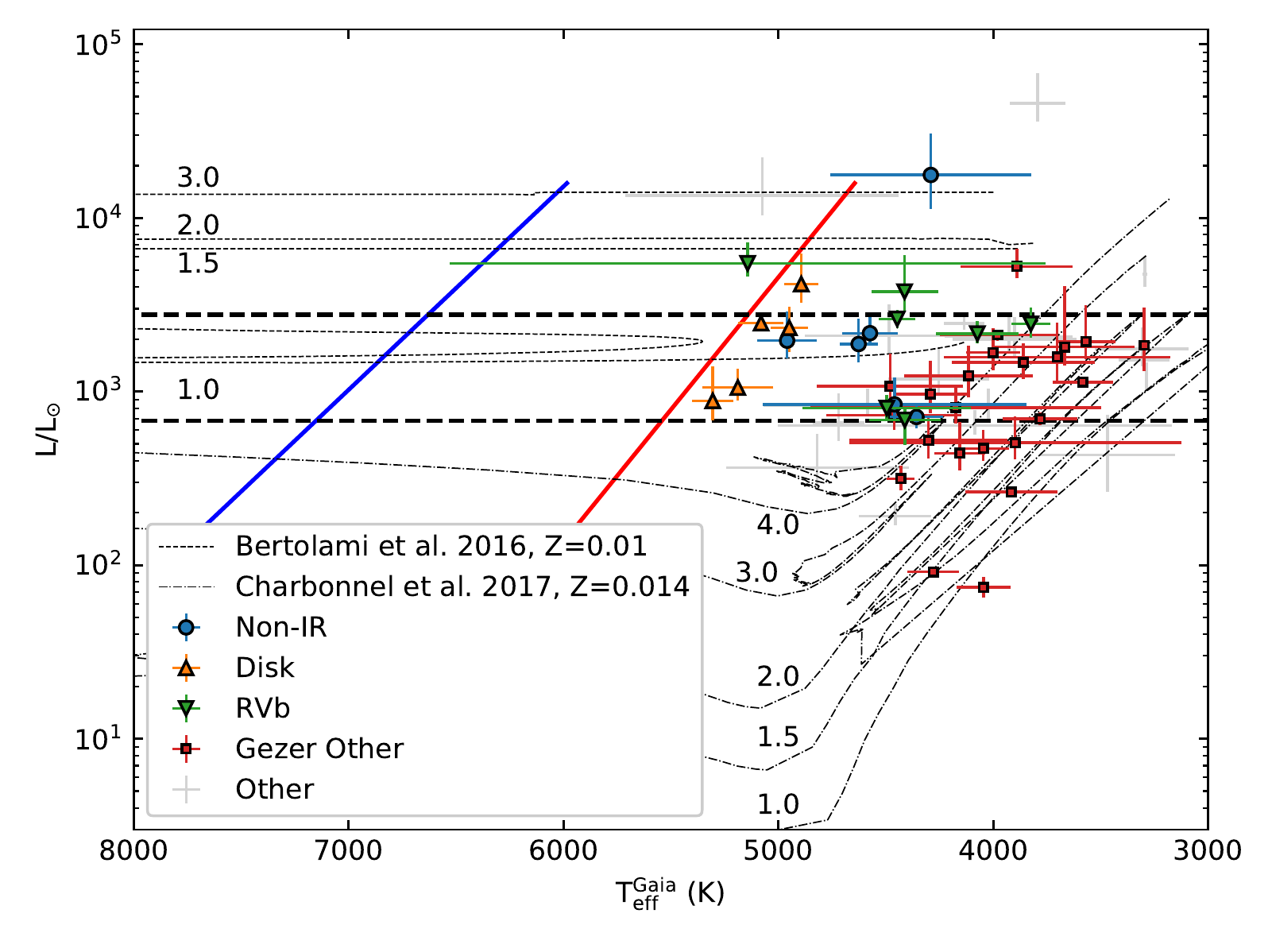}
   \includegraphics[width=9cm]{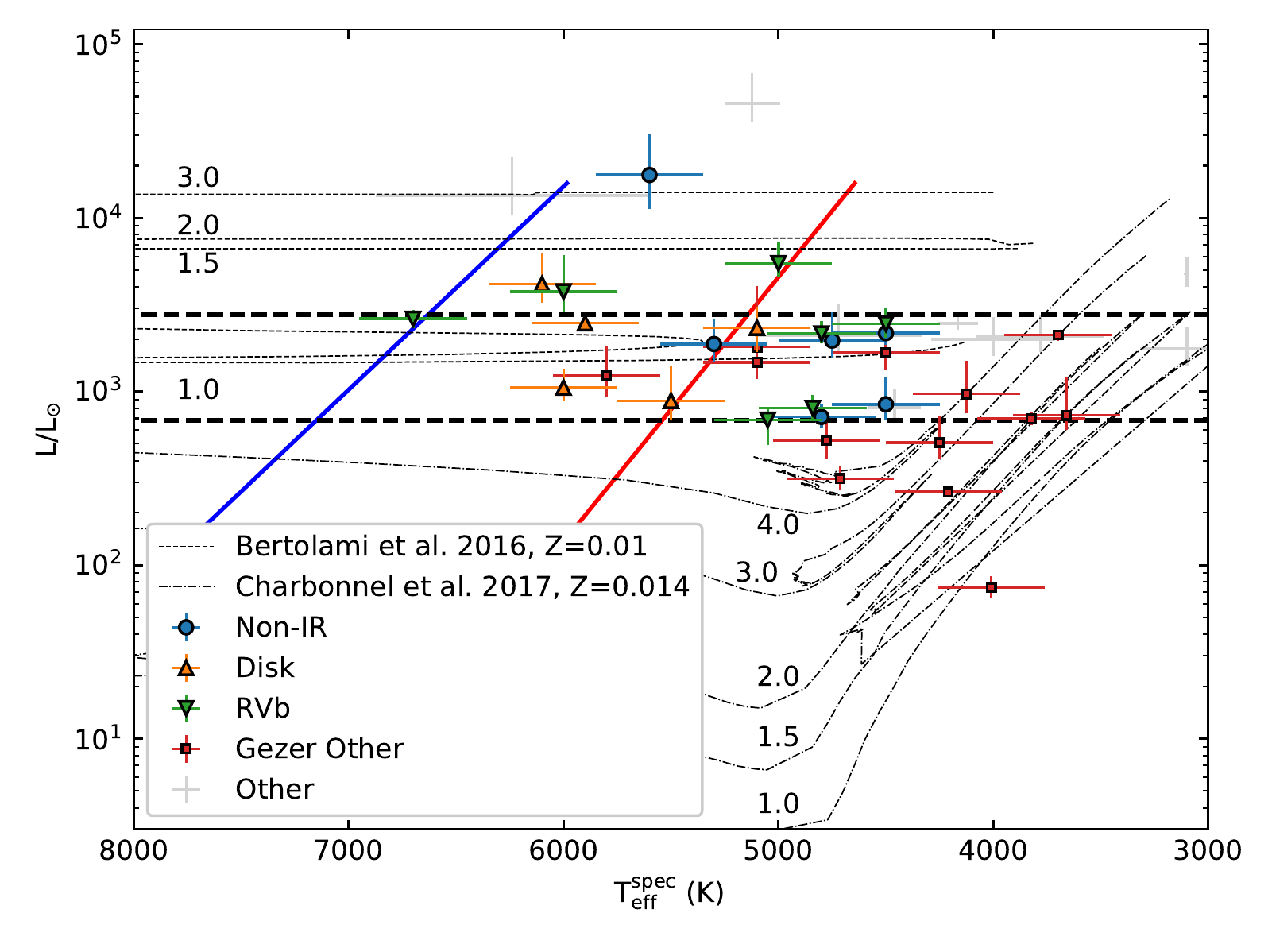}
      \caption{Empirical Hertzsprung--Russell-diagrams of galactic RV Tauri stars. The only difference between the two panels is the choice of effective temperature (upper panel: Gaia DR2; lower panel: spectroscopic temperatures). Blue points: stars that show no sign of IR excess in \citet{Gezer15}; orange triangles: stars with detectable disk in \citet{Gezer15}; green downward triangles: RVb variables; red points: others stars that are also listed in \citet{Gezer15}; grey crosses: other stars from the literature.
      The blue and red diagonal lines indicate the edges of the classical instability strip, adopted from \citet{Christensen03}. The dashed-dotted lines are evolutionary tracks from main sequence to the asymptotic giant branch (AGB), taken from \citet{Charbonnel17}. The black dotted lines are post-AGB evolutionary tracks of \citet{Bertolami16}  with Y=(0.285-0.292). The numbers indicate the initial mass of the models. The black horizontal dashed lines indicate the Tip of the RGB of 1 M$_{\odot}$ (upper) and 4 M$_{\odot}$ (lower) models.}
         \label{HRD}
   \end{figure}

In Fig. \ref{HRD} we show two versions of the empirical Hertzsprung--Russel-diagram. The only difference is the temperature used in the horizontal axes: the top panel is based on the Gaia DR2 temperatures, while the bottom panel was plotted with the spectroscopic effective temperatures. The effect is quite dramatic, given that the Gaia DR2 temperatures only lead to a single RV Tau star falling into the expected instability strip. This clearly indicates that the lack of extinction correction in the Gaia DR2 data makes these temperatures systematically offset. When taking the more reliable spectroscopic temperatures, the majority of the stars is shifted into the instability strip or close to its red edge.

To put both plots in Fig. \ref{HRD} into the context of stellar evolution, we overplotted evolutionary tracks of single low-mass stars from zero-age main sequence to the post-AGB phase \citep{Charbonnel17,Bertolami16}. The blue and the red edges of the classical instability strip were adopted from \citet{Christensen03}. The different symbols of RVa, RVb, dusty and non-dusty stars were used to reveal any dependence of the pulsational characteristics on the presence of a disk. Furthermore, we also highlight the luminosity of the tip of the RGB of 1 M$_{\odot}$ and 4 M$_{\odot}$ models with Z=0.008.

Recently \citet{Manick18} discussed the evolutionary status of SMC and LMC RV Tauri stars, which is based on comparison with single stellar evolutionary models and the relative position of stars to the luminosities of TRGBs. Here we adopt their argumentation regarding the nature of galactic RV Tauri stars.

Looking at the boundaries of the data, it is apparent that disk and RVb stars have luminosities between $\sim$700 L$_{\odot}$ and $\sim$5500 L$_{\odot}$, all falling below the 1.5 M$_{\odot}$ post-AGB track. \citet{Manick18} argued that dusty stars that have higher luminosities than the 1 M$_{\odot}$ TRGB (upper horizontal dash line) are probably post-AGB objects with initial mass higher than $\sim$1 M$_{\odot}$. Stars between the two horizontal lines would be post-AGBs if they are indeed descendants of $\sim$2-4 M$_{\odot}$ stars. Otherwise they were likely formed from lower-mass binary post-RGB progenitors. For lower luminosities, the objects are presumably post-RGB stars with lower progenitor masses. These are probably binaries as all confirmed binary RV Tauri stars are most likely disk sources \citep{Gezer15}, which was recently further strengthened by the RVb analysis of \citet{Kiss17}.

Most of the non-IR galactic RV Tauri stars fall below the theoretical TRGB of an 1 M$_{\odot}$ star, and are near the post-AGB track of an 1 M$_{\odot}$ star. Based on their position in the HRD, they should have gone through a mass-loss phase, but no sign of a dust is detected. \citet{Manick18} speculate that these non-dusty RV Tauri stars are single low-luminosity post-AGB stars with an initial mass lower than 1.25 M$_{\odot}$, where the disk has dispersed on a timescale of 1000 years, making impossible to detect it with recent IR space telescopes (e.g. at 22 micron in case of WISE). However, if they are binaries then the dusty disk has been dispersed during the slow evolution of the low-mass primary.

There is an outlier with much higher luminosity than the others. This star is SS Gem, which is presumably a Pop. I Cepheid as it is well-above the PL relation defined by RV Tauri stars (see Sect. \ref{secPLrel}).

We also note that the location of the several lower luminosity stars with T$_{\rm eff} \leq $5000 K fall close to the blue loops of 3-4 M$_{\odot}$ very metal-poor models (like corresponding to the Magellanic Cloud). However, such massive stars would rather be Pop I Cepheids than Pop II variables, in strong contradiction to the other properties. For example, one of the cool low-luminosity stars is the well-studied DF Cygni, which has a very representative RV Tau nature (see \citealt{Bodi16}).

\citet{Manick18} and \citet{Groenewegen17b} noted that the luminosity of the dusty RV Tauri stars is on average higher, which was attributed by the latter authors to the flux contribution from a companion. The small number of stars in  Fig. \ref{HRD} prevents drawing a similar conclusion, although a slight supporting tendency may be discovered in the distribution of the points (labelled as "Disk" and "RVb" in Fig. \ref{HRD}).  

Overall, the position of the galactic RV Tauri stars in the HR diagram is consistent with those in the Magellanic Clouds. Hence, we can conclude that galactic RV Tauri stars share very similar evolutionary nature despite the different galactic environments.

From a pulsational point of view, all disk stars are located in the theoretical instability strip (IS) within the uncertainties, while a significant fraction of the non-IR and RVb RV Tauris are outside the red edge of the theoretical IS. The fact that some post-AGB stars are located further redward of the IS was already noted by \citet{Kiss07}, but they only found three stars with slightly lower temperatures than expected. Here we can see that this phenomenon is more pronounced, which most likely reflects the structural difference between classical Cepheids and RV Tau stars or the difference between the excitation mechanisms.

\subsection{The period-luminosity relation}
\label{secPLrel}

There is an extensive literature on PL relations of classical pulsating stars, such as RR Lyraes and Cepheids, which we do not  attempt to review here. We only refer to a recent work of \citet{Groenewegen18}, who presented a detailed analysis of PL-relations of Magellanic Cloud Cepheids and related variable stars, including RV Tauris. Our main goal here is to establish 
the first parallax-based PL relation of galactic RV Tau stars.

\begin{figure}
   \centering
   \includegraphics[width=9cm]{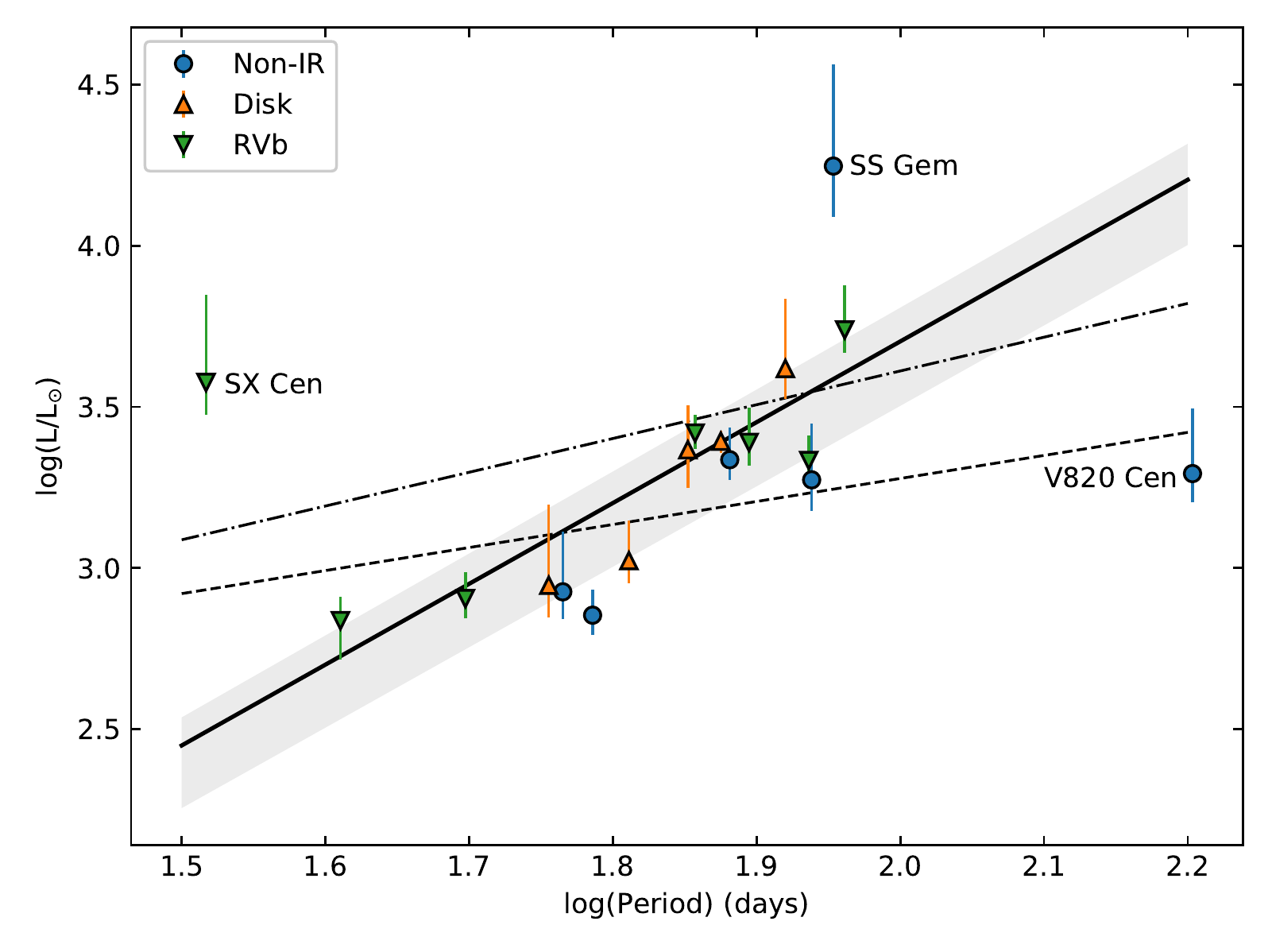}
      \caption{Period-luminosity relation for high confidence galactic RV Tauri type variable stars. The symbols are the same as in Fig. \ref{HRD}. The black line shows the 2$\sigma$ iterative fit to these points for periods smaller than 100 days. The grey shaded area represents the 1$\sigma$ confidence level of the fit. The black dashed and dashed-dotted lines show the PL relation of Population II Cepheids and RV Tauri stars in the Magellanic Clouds, respectively, taken from \citet{Groenewegen17}. The three outliers are SX Cen (RVb), SS Gem (Non-IR) and V820 Cen (Non-IR).}
         \label{PL}
   \end{figure}

Figure \ref{PL} shows the period-luminosity relationship for high-confidence galactic RV Tau type stars, with the data taken from Table \ref{tab1}. There is a noticeable scatter, but most of the points clearly define a linear relationship in the period range of $\sim$40-100 days. However, there is still an outlier, which is the already mentioned overluminous star SS Gem. Moreover, the luminosity of SX Cen and V820 Cen are either too high or to low, respectively, compared to the overall scatter of the relation. 

We fitted a linear function to the logarithmic quantities of these stars with period less than 100 days using an iterative approach with a 2$\sigma$ clipping, which yielded the following equation ($\chi^2 = 0.90$):
\begin{equation}
    {\log L/L_{\odot}} = 2.62^{+0.05}_{-0.01} {\log P} - 1.52^{+0.07}_{-0.22},
\end{equation}
where errors represent the 1$\sigma$ uncertainty and 2 out of the 17 points were excluded. This result is plotted in Fig. \ref{PL} as a black line, where we also depict the inferred PL relation of Population II Cepheids (black dashed line) and RV Tauri stars (black dotted-dashed line) in the Magellanic Clouds of \citet{Groenewegen17}.

Recently, \citet{Groenewegen17} found that RV Tauri-type stars are brighter than expected from the shorter-period Population II Cepheids (BL Her and W Vir objects), i.e. they follow a steeper PL relation. This kind of behaviour for longer period stars has been known for a long time (see e.g. \citet{Harris85}). Here we find an even steeper relation for the galactic RV Tauri stars than in the MCs. \citet{McNamara95} suggested that the reason behind the steeper PL relation is the increase of mass with the period of pulsation. This is, however, in contradiction with the model calculations (see the equations in Sect. \ref{sect_mass}).

Considering the outliers, we searched the literature for any information that could imply that these objects may not be of 
RV Tau-type after all, but we could not find anything conclusive. The physical parameters and the light curve of SX Cen strengthen its evolutionary status. As the PL relation of Pop. I Cepheids lies above that of Pop. II stars, one can naturally conclude that SS Gem may belong to the classical Cepheids instead of RV Tauris. However, its light curve is more similar to those of RV Taus, contradicting the suggestion which comes from the outlying luminosity alone. Finally, V820 Cen follows the PL relation of all Pop. II Cepheids in the 
Magellanic Clouds, hence its position may be a metallicity-related effect (i.e. being more metal-poor than the average in the Milky Way), rather than belonging to a different class of stars.
   
\begin{figure}
   \centering
   \includegraphics[width=9cm]{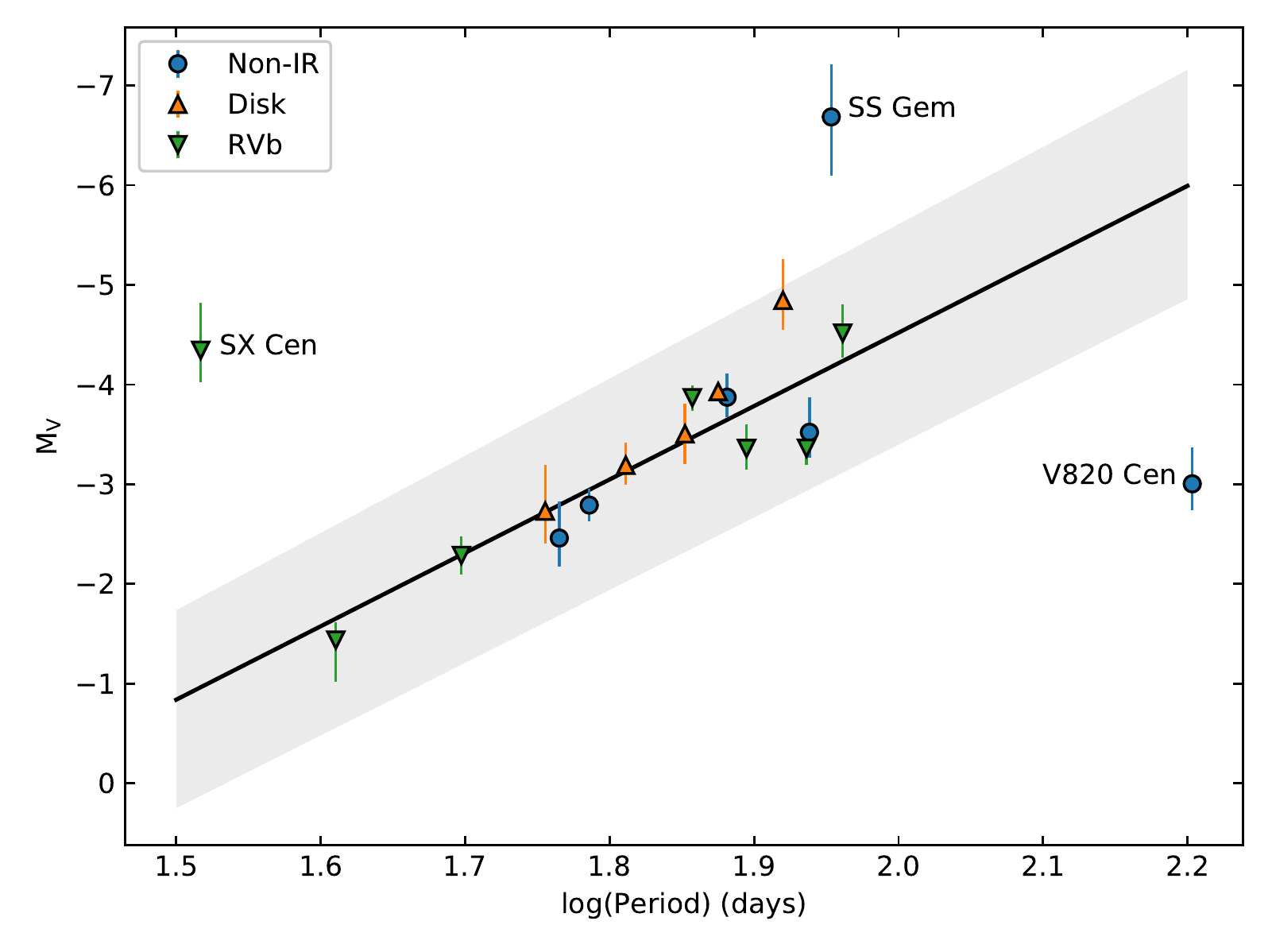}
      \caption{Period-V band absolute magnitude relation. The symbols are the same as in previous figures. The black line shows the fit to these stars where outliers larger than 2$\sigma$ were excluded. The grey shaded area depicts the 1$\sigma$ uncertainty of the fit.}
         \label{PMv}
   \end{figure}

In addition to the period-luminosity relation we have also determined the period-absolute magnitude diagram in the $V$-band in Fig. \ref{PMv}. Here we fitted a linear function to the stars with pulsation periods less than 100 days, which yielded ($\chi^2 = 0.96$):
\begin{equation}
    {\rm M_V} = -7.36^{+0.07}_{-0.38} {\log P} + 10.21^{+0.98}_{-0.33},
\end{equation}
where 3 out of 17 points were excluded. Interestingly, the scatter of the points seems to be smaller and the outliers are the same as in the PL relation. This result can be more easily compared to previous studies because more investigations in the $V$-band are available in the literature.

Some of the previous period-M$_{V}$ relation studies of the Magellanic Cloud and globular cluster variables found that the longer period Pop. II Cepheids follow a steeper slope than BL Her and W Vir stars. The derived slopes are around -4 (-4.35: \citealt{McNamara95}; -3.91: \citealt{Alcock98}; -3.60: \citealt{Harris81}). This mean value is significantly different than ours, which makes our period-M$_{V}$ slope the steepest one ever found. \citet{Soszynski18} published the most recent period vs. Wesenheit index diagram of Pop. II Cepheids of the Magellanic Clouds. Although the authors did not publish any fits,
just the scatter plots in their Fig. 4, a closer look at the data suggests that there is indeed a break in the period-absolute magnitude relation around 20 days. Unfortunately, our sample is too small to draw a firm conclusion and the next Gaia data release will be needed to expand the galactic sample. 

\begin{figure}
   \centering
   \includegraphics[width=9cm]{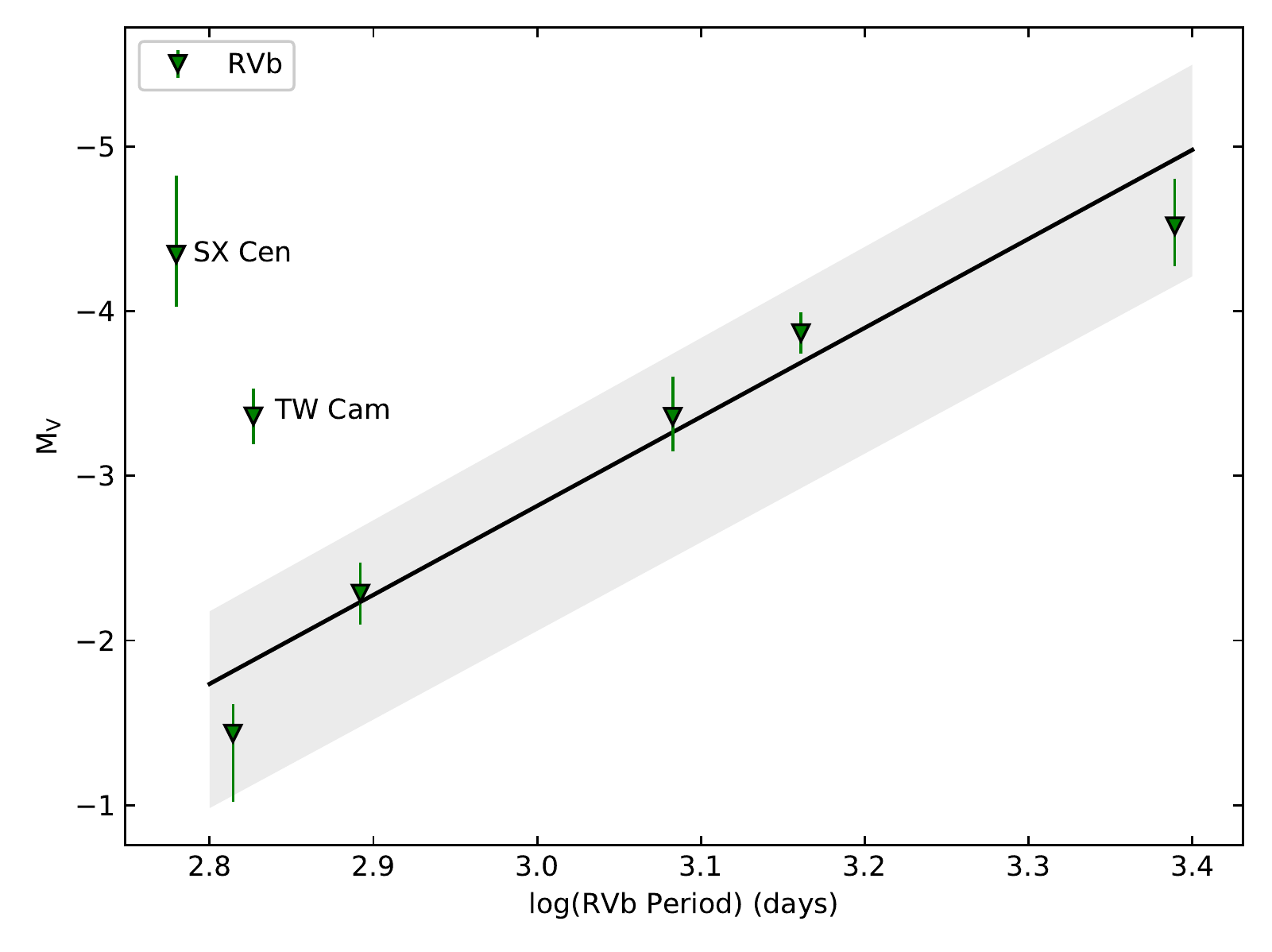}
      \caption{$V$-band absolute magnitude versus period of the mean-brightness variation of RVb stars. The black line shows linear fit to the five points which are in the gray shaded area, which shows the fit uncertainty.}
         \label{PMvRVb}
   \end{figure}

In addition to the traditional period-absolute magnitude relation we noticed an interesting correlation between the RVb period and the absolute magnitude. We plot the V-band absolute magnitude against the period of the slow variation in Fig. \ref{PMvRVb}. We found that 5 out of the 7 RVb stars of our sample follow a strikingly well-determined linear relationship. Interestingly, SX Cen with the shortest period is the same outlier as in the PL relations in Figs. \ref{PL}--\ref{PMv}, the other one is TW Cam. For the sake of completeness we fitted a linear function using the same iterative 2$\sigma$ clipping approach to these points which yielded the following parameters:
\begin{equation}
    {\rm M_V} = -5.403^{+0.023}_{-0.128} {\log P_{\rm RVb} } + 13.390^{+0.689}_{-0.083}.
\end{equation}

\citet{Kiss17} studied the nature of the RVb phenomenon and found supporting evidence for the model of periodic obscuration by a circumbinary dusty disk as an explanation of the slow variations. In this context, the central object is a binary star and the RVb period corresponds to the orbital period of the system. It is not yet clear why there should be an orbital period-absolute magnitude relation for post-AGB binaries, which, if proven, could indicate an important clue about the evolution of these heavily mass-losing binary systems.

\subsection{The period-radius relation}

\citet{Fernie84} collected all available radii of classical Cepheids up to 1982 that were determined using the Baade-Wesselink method and defined a relation between the pulsation period and radius (P--R). This relation was confronted to theoretical expectations \citep{Fernie84,Bono98}, for which an agreement between theory and empirical results have been found for a wide range of periods. \citet{Woolley73} showed that a similar (parallel) relation exists for Pop. II Cepheids (BL Her and W Vir stars). Since then several studies have been conducted on the P--R relation of Type II Cepheids; \citet{Burki86} and \citet{Balog97} presented such a relation in the Milky Way and recently \citet{Groenewegen17} in the Magellanic Clouds.

\begin{figure}
   \centering
   \includegraphics[width=9cm]{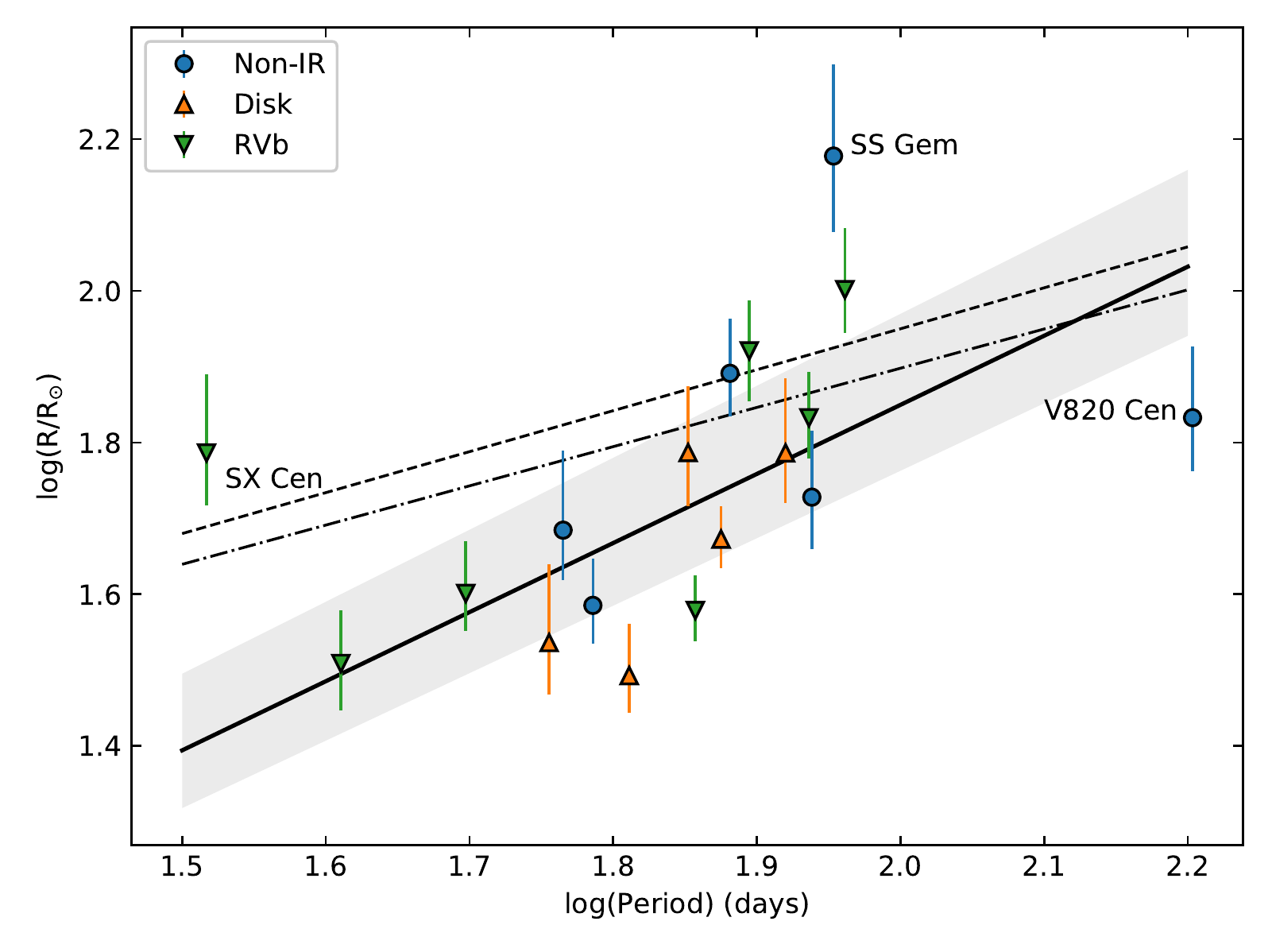}
      \caption{The period-radius relation. The symbols are the same as in previous figures. The black line shows fit to the points. The black dashed and dash-dotted lines indicate the fit of \citet{Burki86} to galactic Pop II Ceps and \citet{Groenewegen17} fit to RV Tauri stars in the Magellanic Clouds, respectively.} 
         \label{Pradius}
   \end{figure}

In Fig. \ref{Pradius} we plot the logarithmic period-radius relation for our sample. The radius dependence on the pulsation period is not so strict, as the points show a relatively large scatter. Nonetheless, a positive correlation is clearly visible, which can be strengthened by an iterative 3$\sigma$ clipping fitting to these points, which yielded the following equation:
\begin{equation}
    {\log R/R_{\odot}} = 0.91^{+0.04}_{-0.02} {\log P} - 0.03^{+0.04}_{-0.04}.
\end{equation}
This fit can be seen as a black line with the 1$\sigma$ confidence level in Fig. \ref{Pradius}.

The labeled outliers are the same as in the previous PL plots. As the radii were calculated from the luminosities and effective temperatures, it is not surprising that SS Gem have the largest radius. Within the given errors, the radius of V820 Cen just happen to follow the fitted relationship. However, if we force to exclude this star in the fitting process, we get a slightly steeper slope.

The P--R relation of BL Her and W Vir stars of \citet{Burki86} and of RV Tauris in the Magellanic Clouds by \citet{Groenewegen17} are shown by black dashed and dash-dotted lines in Fig. \ref{Pradius}, respectively. \citet{Balog97} did not publish their results quantitatively, so we cannot directly compare theirs to ours. As can be seen in Fig. \ref{Pradius}, the PR relation of \citet{Burki86} and \citet{Groenewegen17} lie close to each other. Contrary to this, our sample of galactic RV Tauri stars appear to follow a steeper relation with a deviation larger than the uncertainties. As the plotted linear of \citet{Burki86} is only an extrapolation of a fit to BL Her and W Wir stars, the deviation may arise from the different types of Pop. II Cepheids, even keeping in mind that their study is also based on a sample of galactic stars. The fit of \citet{Groenewegen17} covers the RV Tau regime, which makes it challenging to explain the observed deviation. As the metallicity is the main difference between the Milky Way and the Magellanic Clouds, a natural explanation could be a [Fe/H]-dependent PR relation; \citet{Groenewegen17} did not find any sign of it, hence this issue is also in need of further data in the next Gaia data release.

\subsection{Mass estimates}
\label{sect_mass}

\citet{Marconi15} computed a large grid of nonlinear, time-dependent convective hydrodynamical models of fundamental and first overtone pulsators assuming a broad range in metal abundances (Z = 0.0001--0.02). Based on these models they constructed new metal-dependent pulsation relations, i.e. the correlations between pulsation and evolutionary observables. They omitted the most luminous models (called Sequence D; L/L$_\odot\sim 100$), because such high values are untypical in case of RR Lyraes, the main targets of their study. However, Type II Cepheids lie in this higher luminosity range. \citet{Groenewegen17} re-derived these relations considering all models with log$L > 1.65 L_\odot$ and found the following equation for fundamental mode pulsators:
\begin{equation} \label{mass_eq1}
\begin{split}
    \log P =&\  (11.468\pm0.049) 
    + (0.8627\pm0.0028)\log L \\
    &-(0.617\pm0.015)\log M
    -(3.463\pm0.012)\log T_{\rm eff} \\
    &+ (0.0207\pm0.0013)\log Z \
    (N=195,\sigma=0.0044). \\
\end{split}
\end{equation}

Previously, \citet{Bono00} computed a set of nonlinear, convective Cepheid models. They found that for each metallicity, the predicted fundamental periods can be connected to mass, luminosity, and effective temperature. \citet{Groenewegen17} combined the "canonical" and "non-canonical" models of \citet{Bono00} and found for fundamental mode pulsators:
\begin{equation} \label{mass_eq2}
\begin{split}
    \log P =&\  (10.649\pm0.085) + 
    (0.9325\pm0.0053)\log L \\
    &-(0.799\pm0.020)\log M
    -(3.282\pm0.022)\log T_{\rm eff} \\
    &+ (0.0393\pm0.0026)\log Z \ 
    (N = 202, \sigma = 0.0085). \\
\end{split}
\end{equation}

If we know the pulsation period, luminosity, effective temperature and metallicity, these equations can be used to estimate the stellar mass. \citet{Groenewegen17} tested the method on a known classical Cepheid (OGLE-LMC-CEP-0227) and found agreement with the literature within the error bars. To estimate masses using Eq. \ref{mass_eq1} and \ref{mass_eq2} we assumed Z=0.014 (solar metallicity). The results are listed in Table \ref{tab:masses}.

\begin{table}
\centering
\caption{The estimated masses of high-confidence galactic RV Tauri stars. The RRL and Cep subscripts refer to the papers of \citet{Marconi15} and \citet{Bono00}, respectively, which were used for the calculations (see text for details). The errors were estimated from the uncertainties of the physical parameters.}
\begin{tabular}{ccc}
\hline \hline
Name & M$_{\rm RRL}$/M$_{\odot}$ & M$_{\rm Cep}$/M$_{\odot}$\\
\hline
\multicolumn{3}{c}{RVa type stars} \\
\hline
UY Ara & 0.20$^{+0.17}_{-0.08}$ & 0.31$^{+0.22}_{-0.10}$\\
EQ Cas & 0.55$^{+0.37}_{-0.23}$ & 0.65$^{+0.36}_{-0.21}$\\
RU Cen & 0.13$^{+0.06}_{-0.04}$ & 0.23$^{+0.09}_{-0.06}$\\
V820 Cen & 0.26$^{+0.18}_{-0.11}$ & 0.40$^{+0.23}_{-0.13}$\\
SS Gem & 5.60$^{+5.84}_{-3.16}$ & 5.37$^{+4.64}_{-2.47}$\\
AC Her & 0.35$^{+0.09}_{-0.09}$ & 0.54$^{+0.11}_{-0.11}$\\
EP Lyr & 0.52$^{+0.38}_{-0.20}$ & 0.77$^{+0.47}_{-0.23}$\\
TT Oph & 0.28$^{+0.11}_{-0.10}$ & 0.39$^{+0.12}_{-0.10}$\\
R Sge & 0.81$^{+0.43}_{-0.38}$ & 0.99$^{+0.42}_{-0.37}$\\
AR Sgr & 0.35$^{+0.22}_{-0.14}$ & 0.51$^{+0.26}_{-0.16}$\\
V Vul & 1.33$^{+0.60}_{-0.49}$ & 1.40$^{+0.50}_{-0.40}$\\
\hline
\multicolumn{3}{c}{RVb type stars} \\
\hline
TW Cam & 0.75$^{+0.29}_{-0.25}$ & 0.91$^{+0.27}_{-0.23}$\\
IW Car & 0.20$^{+0.06}_{-0.05}$ & 0.37$^{+0.08}_{-0.07}$\\
SX Cen & 2.23$^{+2.03}_{-0.88}$ & 2.34$^{+1.76}_{-0.74}$\\
DF Cyg & 0.45$^{+0.18}_{-0.16}$ & 0.56$^{+0.17}_{-0.15}$\\
BT Lac & 0.38$^{+0.14}_{-0.18}$ & 0.50$^{+0.14}_{-0.19}$\\
U Mon & 2.00$^{+1.06}_{-0.72}$ & 2.13$^{+0.91}_{-0.59}$\\
RV Tau & 1.50$^{+0.70}_{-0.58}$ & 1.56$^{+0.57}_{-0.46}$\\
\hline
\end{tabular}
\label{tab:masses}
\end{table}

The estimated masses are in the range of $\sim$0.1-2.2 M$_{\odot}$ independently from the period. There is only one outlier with significantly higher value, SS Gem. The resulting masses of the two methods are in agreement within the given errors. They differ each other mostly by $0.1-0.2$ M$_\odot$. If we take the mean of the two kind of masses and separate the non-IR, disk and RVb stars, we find the following. The masses of the non-IR RV Tauri stars are in the range of 0.33-5.48 M$_{\odot}$ with a median of 0.52 M$_{\odot}$, the masses of the disk stars are in the range of 0.18-0.9 M$_{\odot}$ with a median of 0.45 M$_{\odot}$, while the mass of the RVb stars is in the range of 0.28-2.28 M$_{\odot}$ with a median of 0.83 M$_{\odot}$. We note that these values are based on half of the formal periods, because the usage of double periods resulted in unphysically low masses. This phenomenon may imply that the real pulsation period of RV Tauri stars is the elapsed time between two consecutive minima aside from its depth.

   \begin{figure}
   \centering
   \includegraphics[width=9cm]{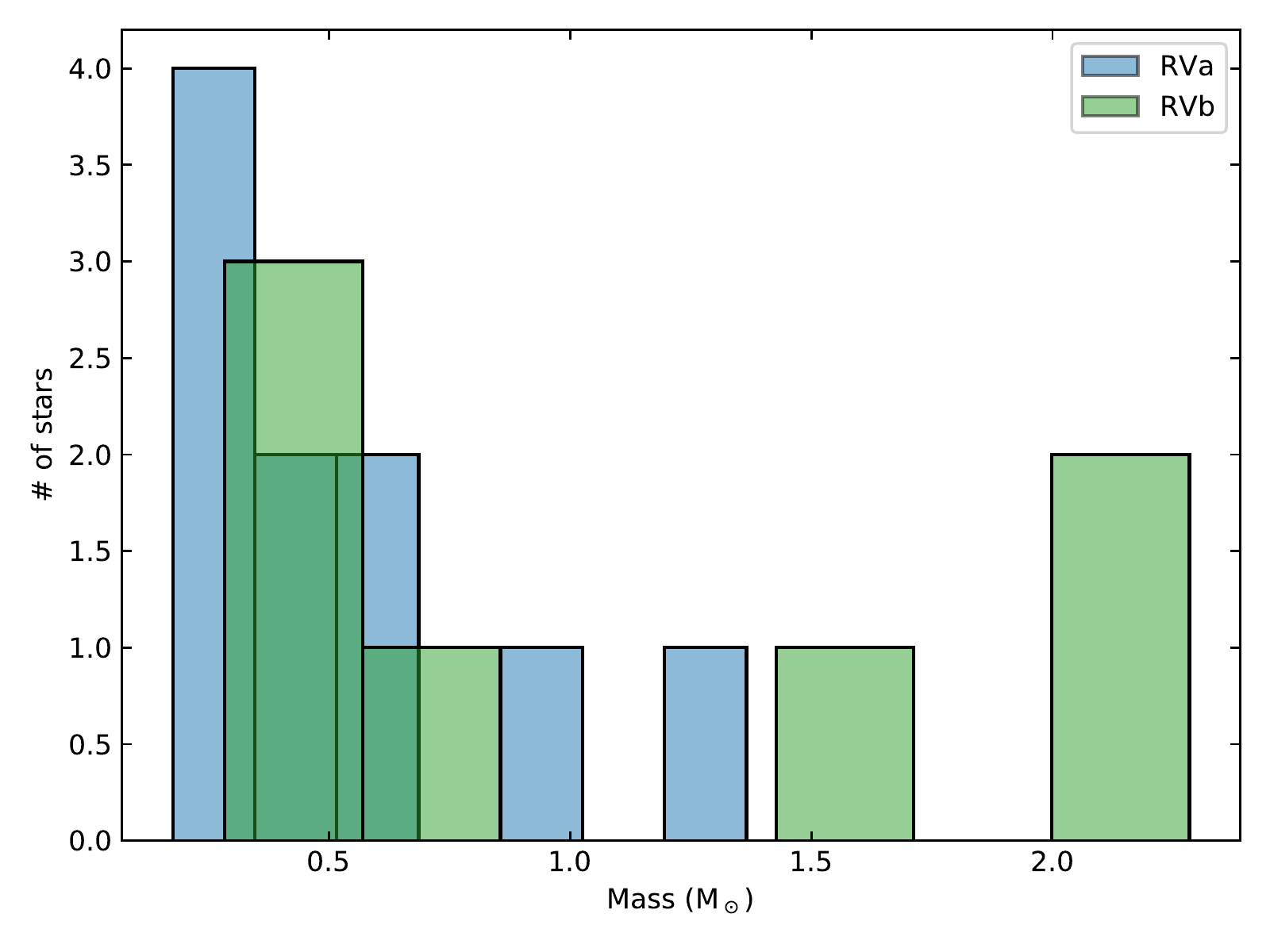}
      \caption{The distribution of estimated masses of high-confidence galactic RV Tauri stars. Note that SS Gem with the highest mass (above 5 M$_\odot$) is located way outside the shown range in the horizontal axis.} 
         \label{Fig:masses}
   \end{figure}

The estimated masses spread on a wide range (see. Fig \ref{Fig:masses}). Our results are generally consistent with those of \citet{Groenewegen17} in the Magellanic Clouds. If we look at the median masses of the different types of RV Tau stars, we can recognize the significant difference between non-IR/IR and RVb stars; non-IR and dusty ones have similar, median values (0.45--0.52 M$_{\odot}$), while the RVb stars have approximately the double of that (0.83 M$_{\odot}$). However, it is important to note that the RVb masses follow a bimodal distribution ($\sim$0.7 M$_{\odot}$ and $\sim$1.8 M$_{\odot}$), which prevents drawing a firm conclusion. Just as above, the sample size is critical here and further stars will be needed when the distance limit of the Gaia data will be pushed further away.

Stars with masses greater than 1 M$_{\odot}$ nearly follow the expectations from their position in the HRD compared to the single post-AGB theoretical evolutionary tracks, as the relevant models have initial masses of 0.8--1.5 M$_{\odot}$. However, this comparison would be more relevant if we were using binary model calculations (see Sect. \ref{sect:HRD}). Regarding the lower mass stars, those ones that have masses around 0.5--0.6 M$_{\odot}$ agree with the model calculations of fundamental pulsators of \citet{Bono97}. All put together, we find that the derived physical parameters are more or less consistent with the theoretical expectations. 

Finally, we note that the only star with significantly higher mass is SS Gem, with a mass of $\sim$5.48 M$_{\odot}$. Such a large value is again typical for Pop. I Cepheids \citep{Turner96}, which gives another supporting evidence to the previous conclusion that SS Gem is likely to be a massive young supergiant star instead of a post-AGB pulsator.

\section{Summary}

We have compiled a carefully selected list of galactic RV Tauri stars. We took the dominant period values from the literature or determined by ourselves when needed from the available light curve data. Then we cross matched our list of coordinates with the Gaia DR2 database. To infer distances, bolometric magnitudes, luminosities, and radii, we used the slightly modified version of {\tt isoclassify} code of \citet{Huber17}, which uses a Monte-Carlo sampling scheme and derives posterior distributions.

As the evolutionary status of several objects have been questioned in the literature or is uncertain, we restricted our sample to well-studied stars. To do so, we used the chemically studied sample of \citet{Gezer15} and the RVb variables of \citet{Kiss17} as our high-confidence sample. To carry out our analysis, we created the most reliable, high-confidence collection of galactic RV Tauri-type variables with well-determined Gaia DR2 distances, which contains 12 RVa- and 6 RVb-type stars in the Galaxy.

The main work in this paper is that we derived parallax-based period-luminosity and period-radius relations for galactic RV Tauri-type variable stars. The most important results inferred from our analysis can be listed as follows:
\begin{enumerate}
    \item We showed that Gaia DR2 effective temperatures for RV Tau-tye stars deviate significantly from the spectroscopically determined values. They are lower with a median shift of $\sim$436 K. The reason for this systematics is the lack of reddening correction for stars that lie in the location of RV Tau-type stars in the Hertzsprung--Russell-diagram. 
    
    \item We discussed the evolutionary status of galactic RV Tau-type stars, which is fairly ambiguous. The most luminous ones that are brighter than the TRGB of 1 M$_{\odot}$ model are presumably post-AGB objects that are descendants of stars with masses higher than 1 M$_\odot$. Fainter ones are probably post-AGBs if they have an initial mass between $\sim$2-4 M$_{\odot}$. Otherwise they likely were formed from lower mass binary post-RGB progenitors. Others are post-RGB binary stars with lower progenitor masses.
    
    \item From the position of stars in the HR diagram we conclude that the instability strip of RV Tauri stars has a broader extension in the cooler range than the classical IS of 
    classical Cepheids.
    
    \item The galactic RV Tauris follow steeper period-luminosity and period-radius relations than those of the Population II Cepheids with shorter pulsation periods. 
    
    \item For the first time, we derived a period-absolute magnitude relation between the period of the mean-brightness variation of RVb stars and their $V$-band absolute magnitude. However, this relation is based on a very low number of stars; further observations will be needed to confirm this correlation.
    
    \item We found that the median mass of RVa stars is around 0.45--0.52 M$_{\odot}$, which is in agreement with Type II Cepheid model calculations. The mass distribution of our very small sample of RVb stars is sort of bimodal, with masses around $\sim$0.7 M$_{\odot}$ and $\sim$1.8 M$_{\odot}$.
    
\end{enumerate}

Further understanding of galactic RV Tau-type stars will need the more accurate next Gaia data release, which is expected to increase the sample size significantly. 

\acknowledgments

This work has been supported by the Lend\"ulet LP2018-7/2018, the NKFIH K-115709 and the GINOP-2.3.2-15-2016-00003 grants of the Hungarian National Research, Development and Innovation Office, and the Hungarian Academy of Sciences. This research has made use of the International Variable Star Index (VSX) database, operated at AAVSO, Cambridge, Massachusetts, USA. We acknowledge with thanks the variable star observations from the AAVSO International Database contributed by observers worldwide and used in this research. This research has made use of the SIMBAD database,
operated at CDS, Strasbourg, France.

\clearpage

\begin{appendix}
\setcounter{table}{0}
\renewcommand{\thetable}{A\arabic{table}}

In the Appendix we list the physical parameters of stars that have not been discarded by our initial criteria, but were not used in the analysis, because several of them have luminosities that are much lower than expected from post-AGB evolution.

\begin{table*}[!ht]
\caption{Derived physical parameters of other galactic RV Tau stars that are listed in \citet{Gezer15}. The errors represent the 1$\sigma$ confidence level of the posterior distributions. The effective temperatures were taken from the literature. Pulsation periods were determined by us or were taken from the literature.}
\centering
\begin{tabular}{cccccccccc}
\hline \hline
Name & d [pc] & L/L$_{\odot}$ & R/R$_{\odot}$ & M$_{\rm V}$ [mag] & T$_{\rm eff,lit}$ [K] & Period [d] & Ref. & $\pi$ [mas] & $\sigma_\pi$ [mas]\\
\hline
DY Aql & 1778$^{+214}_{-161}$ & 508$^{+213}_{-102}$ & 38.7$^{+8.4}_{-6.6}$ & -1.857$^{+0.310}_{-0.347}$ & 4250 & 131.86 & 3 & 0.57 & 0.06 \\
V381 Aql & 3146$^{+467}_{-334}$ & 809$^{+281}_{-153}$ & \nodata & \nodata & \nodata & 109.60 & 3 & 0.32 & 0.04 \\
V686 Ara & 2064$^{+171}_{-122}$ & 74$^{+12}_{-10}$ & 17.8$^{+3.0}_{-2.4}$ & 0.208$^{+0.149}_{-0.170}$ & 4010 & 36.30 & 3 & 0.48 & 0.03 \\
V662 Ara & 3476$^{+890}_{-445}$ & 1579$^{+926}_{-421}$ & \nodata & -1.564$^{+0.330}_{-0.471}$ & \nodata & 92.80 & 3 & 0.28 & 0.05 \\
V428 Aur & 585$^{+21}_{-19}$ & 2118$^{+156}_{-156}$ & 120.2$^{+20.0}_{-15.0}$ & -2.431$^{+0.076}_{-0.076}$ & 3699 & 110.57 & 2 & 1.71 & 0.06 \\
GK Car & 4189$^{+733}_{-440}$ & 729$^{+478}_{-130}$ & 95.1$^{+28.0}_{-18.7}$ & -1.743$^{+0.383}_{-0.485}$ & 3660 & 55.30 & 3 & 0.23 & 0.03 \\
V1071 Cas & 1561$^{+55}_{-48}$ & 91$^{+6}_{-7}$ & \nodata & 1.124$^{+0.081}_{-0.081}$ & \nodata & 96.70 & 3 & 0.64 & 0.02 \\
V345 Cen & 2549$^{+727}_{-363}$ & 1848$^{+1199}_{-533}$ & \nodata & \nodata & \nodata & 166.80 & 3 & 0.39 & 0.07 \\
BU Cen & 3458$^{+564}_{-376}$ & 1678$^{+640}_{-349}$ & 61.6$^{+12.7}_{-10.1}$ & -1.094$^{+0.267}_{-0.320}$ & 4500 & 170.40 & 3 & 0.29 & 0.04 \\
SY Cir & 5151$^{+817}_{-613}$ & 523$^{+203}_{-113}$ & 31.7$^{+6.7}_{-4.6}$ & \nodata & 4777 & 46.36 & 3 & 0.19 & 0.03 \\
V399 Cyg & 4073$^{+526}_{-394}$ & 1473$^{+417}_{-292}$ & 41.6$^{+7.5}_{-6.0}$ & -2.382$^{+0.243}_{-0.292}$ & 5100 & 142.45 & 2 & 0.24 & 0.03 \\
V457 Cyg & 5066$^{+1108}_{-665}$ & 1232$^{+610}_{-305}$ & 31.3$^{+7.2}_{-4.7}$ & \nodata & 5800 & 80.35 & 2 & 0.19 & 0.03 \\
CW Ind & 3121$^{+287}_{-205}$ & 315$^{+58}_{-45}$ & 24.9$^{+3.9}_{-3.3}$ & -1.081$^{+0.167}_{-0.167}$ & 4713 & 121.21 & 3 & 0.32 & 0.02 \\
V338 Lib & 1694$^{+194}_{-138}$ & 470$^{+130}_{-72}$ & \nodata & -0.716$^{+0.177}_{-0.236}$ & \nodata & 310.00 & 3 & 0.59 & 0.06 \\
V338 Mus & 1021$^{+38}_{-35}$ & 695$^{+65}_{-54}$ & 58.2$^{+9.6}_{-6.8}$ & -0.823$^{+0.095}_{-0.095}$ & 3824 & 206.70 & 3 & 0.98 & 0.03 \\
V407 Pav & 5319$^{+1111}_{-793}$ & 968$^{+542}_{-217}$ & 67.5$^{+18.4}_{-11.5}$ & -2.389$^{+0.306}_{-0.408}$ & 4126 & 112.80 & 3 & 0.18 & 0.03 \\
V360 Peg & 614$^{+22}_{-21}$ & 1133$^{+85}_{-85}$ & \nodata & -0.970$^{+0.077}_{-0.077}$ & \nodata & 90.53 & 2 & 1.63 & 0.06 \\
V894 Per & 2126$^{+195}_{-140}$ & 5257$^{+1390}_{-772}$ & \nodata & -4.139$^{+0.209}_{-0.261}$ & \nodata & 69.70 & 4 & 0.47 & 0.04 \\
V594 Pup & 5721$^{+1240}_{-767}$ & 1071$^{+581}_{-274}$ & \nodata & -3.091$^{+0.290}_{-0.435}$ & \nodata & 57.86 & 1 & 0.16 & 0.03 \\
V760-Sgr & 2145$^{+339}_{-226}$ & 1806$^{+2240}_{-395}$ & 48.4$^{+23.3}_{-7.1}$ & -3.431$^{+0.233}_{-1.116}$ & 5100 & 44.98 & 1 & 0.47 & 0.06 \\
V1284 Sgr & 3355$^{+753}_{-376}$ & 441$^{+224}_{-89}$ & \nodata & \nodata & \nodata & 80.03 & 1 & 0.29 & 0.05 \\
HD 172810 & 434$^{+10}_{-9}$ & 264$^{+15}_{-13}$ & 27.5$^{+3.9}_{-3.3}$ & -0.080$^{+0.054}_{-0.061}$ & 4210 & \nodata & \nodata & 2.30 & 0.05 \\
\hline
\end{tabular}
\label{tab3}
\tablerefs{ (1) \citet{Wils08}; (2) this paper; (3) \citet{Samus17}; (4) \citet{Kazarovets11} }
\end{table*}

\begin{table*}[!ht]
\caption{Derived physical parameters of other RV Tau stars that were observed by Gaia. The errors represents the 1$\sigma$ confidence level of the posterior distributions. The effective temperatures were taken from the literature. Pulsation periods were determined by us or were taken from the literature, respectively.}
\centering
\begin{tabular}{cccccccccc}
\hline \hline
Name & d [pc] & L/L$_{\odot}$ & R/R$_{\odot}$ & M$_{\rm V}$ [mag] & T$_{\rm eff,lit}$ [K] & Period [d] & Ref. & $\pi$ [mas] & $\sigma_\pi$ [mas]\\
\hline
KK Aql & 2883$^{+357}_{-268}$ & 2074$^{+690}_{-384}$ & 109.7$^{+24.0}_{-18.0}$ & -2.887$^{+0.240}_{-0.288}$ & 3780 & 106.00 & 3 & 0.35 & 0.04 \\
RY Ara & 2603$^{+549}_{-310}$ & 2101$^{+1061}_{-424}$ & 68.2$^{+17.1}_{-10.9}$ & \nodata& 4720 & 144.00 & 1 & 0.38 & 0.06 \\
AG Aur & 2437$^{+421}_{-230}$ & 1995$^{+691}_{-395}$ & 93.1$^{+21.7}_{-13.6}$ & -2.651$^{+0.230}_{-0.323}$ & 4000 & 96.00 & 3 & 0.40 & 0.05 \\
UZ CMa & 955$^{+93}_{-75}$ & 1193$^{+277}_{-185}$ & 289.8$^{+69.9}_{-54.4}$ & 1.045$^{+0.195}_{-0.207}$ & 2800 & 362.00 & 3 & 1.06 & 0.09 \\
DI Car & 3543$^{+359}_{-269}$ & 834$^{+212}_{-106}$ & 11.5$^{+1.4}_{-1.1}$ & -2.437$^{+0.179}_{-0.215}$ & 9400 & 58.31 & 2 & 0.28 & 0.03 \\
PY Cas & 1092$^{+108}_{-77}$ & 4755$^{+1227}_{-767}$ & 309.0$^{+72.7}_{-45.4}$ & -1.742$^{+0.215}_{-0.259}$ & 3100 & 110.47 & 2 & 0.92 & 0.08 \\
V405 Cen & 2029$^{+135}_{-102}$ & 638$^{+144}_{-72}$ & \nodata & \nodata& \nodata & 33.80 & 3 & 0.49 & 0.03 \\
SX Her & 1532$^{+68}_{-61}$ & 2471$^{+236}_{-207}$ & 93.7$^{+14.1}_{-10.1}$ & -2.513$^{+0.095}_{-0.095}$ & 4165 & 102.90 & 3 & 0.65 & 0.03 \\
VV Mus & 8381$^{+985}_{-985}$ & 1178$^{+586}_{-204}$ & \nodata & \nodata& \nodata & 29.00 & 3 & 0.10 & 0.02 \\
V564 Oph & 1777$^{+142}_{-118}$ & 807$^{+234}_{-128}$ & 44.3$^{+8.1}_{-6.3}$ & -1.685$^{+0.215}_{-0.269}$ & 4460 & 70.33 & 3 & 0.57 & 0.04 \\
EI Peg & 816$^{+70}_{-54}$ & 1522$^{+279}_{-175}$ & \nodata & -0.620$^{+0.152}_{-0.172}$ & \nodata & 61.15 & 3 & 1.24 & 0.09 \\
GK Sct & 3182$^{+794}_{-433}$ & 432$^{+300}_{-167}$ & \nodata & \nodata& \nodata & 40.00 & 3 & 0.32 & 0.06 \\
RS Sge & 5563$^{+1125}_{-750}$ & 364$^{+206}_{-94}$ & \nodata & \nodata& \nodata & 82.39 & 3 & 0.17 & 0.03 \\
S Sge & 1601$^{+329}_{-197}$ & 13490$^{+8766}_{-3188}$ & 99.5$^{+26.0}_{-14.5}$ & -6.145$^{+0.252}_{-0.554}$ & 6240 & 16.76 & 2 & 0.64 & 0.09 \\
DZ UMa & 3127$^{+764}_{-424}$ & 1940$^{+1209}_{-403}$ & \nodata & \nodata& \nodata & 140.94 & 2 & 0.32 & 0.06 \\
S Vul & 3249$^{+528}_{-352}$ & 45690$^{+22810}_{-9777}$ & 240.9$^{+62.0}_{-35.4}$ & -6.737$^{+0.337}_{-0.433}$ & 5123 & 68.46 & 3 & 0.30 & 0.04 \\
TYC 3229-1483-1 & 1951$^{+137}_{-114}$ & 192$^{+33}_{-22}$ & \nodata & -0.420$^{+0.125}_{-0.166}$ & \nodata & 44.60 & 4 & 0.52 & 0.03 \\
\hline
\end{tabular}
\tablerefs{ (1) \citet{Wils08}; (2) this paper; (3) \citet{Samus17}; (4) \citet{Dimitrov07} }
\end{table*}

\end{appendix}

\end{document}